\documentstyle[epsf,twocolumn,aps]{revtex}
\input{epsf}
\begin{document}
\draft
\addtolength\textheight{40pt}
\addtolength\footskip{-10pt}
\def\authornote#1{$\bullet$ {\sc #1}}

\twocolumn[\hsize\textwidth\columnwidth\hsize\csname @twocolumnfalse\endcsname

\title{Pseudogap above $T_c$ in a model with ${d_{x^2-y^2}}$ pairing}

\author{Jan R.~Engelbrecht and Alexander Nazarenko}
\address{Department of Physics, Boston College, Chestnut Hill, MA 02167}
\author{Mohit Randeria}
\address{Theoretical Physics Group, Tata Institute of Fundamental Research,
Mumbai 400005, India}
\author{Elbio Dagotto}
\address{Department of Physics and National High Magnetic Field Lab,
Florida State University, Tallahassee, FL 32306}
\maketitle

\begin{abstract}

We study the anomalous normal state properties
of a simple two-dimensional model whose ground state is a 
$d$-wave superconductor. Using a self-consistent, conserving
formulation, we show that pairing correlations above $T_c$ 
lead to the appearance of a highly anisotropic pseudogap
in the electronic spectral function and the destruction of the
Fermi surface. We discuss the similarities and differences
between our results and ARPES experiments on underdoped cuprates.

\end{abstract}


\vskip2pc]
\narrowtext

The deviations from Fermi liquid theory (FLT) in the normal state of
high $T_c$ superconductors are now well established. 
It has recently become clear
that the underdoped cuprates exhibit even more remarkable deviations from
FLT than the optimally doped materials: not only are the quasiparticles
not defined, but the Fermi surface also becomes ill-defined due to
the opening of a pseudogap. Early evidence for the suppression of low
frequency spectral weight above $T_c$ came from a variety of probes, including
NMR\cite{NMR}, specific heat\cite{SP-HEAT} and optics\cite{OPTICS}.
Recent angle-resolved photoemission (ARPES) studies\cite{STANFORD,ARGONNE}
on Bi-2212 compounds
have considerably clarified the situation by providing direct evidence
for a highly anisotropic pseudogap, for $T_c < T < T^*$, which is similar
in its magnitude and in its angular dependence to the d-wave
superconducting gap below $T_c$.

These observations of a normal state pseudogap find a simple explanation
in a theory\cite{CROSSOVER,NEGU} in which the pairing amplitude develops 
at a crossover scale $T^*$ higher than the 
$T_c \sim n_s/m^*$ \cite{UEMURA,EMERY} at which phase coherence sets in.
The separation between $T_c$ and $T^*$ naturally occurs in low density,
short coherence length superconductors, and leads to striking deviations
from FLT in degenerate Fermi systems in 2D\cite{NEGU}. However, rather little
is known theoretically about the pseudogap state above $T_c$
in d-wave superconductors
(since, for technical reasons, the quantum Monte Carlo results\cite{NEGU}
are restricted to s-wave pairing). Such a study is clearly important,
not only because the experiments show  d-wave pairing, but also
to compare the predictions of these theories\cite{CROSSOVER,NEGU,EMERY} with
those of RVB-based theories\cite{RVB} in which spinons pair at $T^*$ and 
holons condense at $T_c$.  An improved understanding of the origin of this
pseudogap effect may also be an important clue towards a theory of high
temperature superconductivity.

As a first step in this direction, we study a phenomenological model
of a 2D d-wave superconductor, in a parameter range where the 
normal state is dominated by d-wave pairing correlations.
Using a self-consistent, conserving aproximation, we show that there
is a crossover temperature scale $T^*$ below which
normal state spectral functions exhibit anomalous dispersion  
with a highly anisotropic suppression 
of spectral weight -- pseudogap -- near
the chemical potential. This also leads to the  partial destruction of 
the Fermi surface along certain directions in the Brillouin
zone. Both these effects are very similar to the ARPES experiments.
We also find that the normal state spin susceptibility acquires
a spin-gap like $T$-dependence. 

Let us consider a simple two-dimensional model which has a 
superconducting ground state with $d_{x^2-y^2}$ symmetry, defined by
\begin{equation}
\label{eq:H}
H\!=\!\sum_{{\bf{k}},\sigma}(\epsilon_{\bf{k}}\!-\!\mu)
c^{\dagger}_{{\bf{k}}\sigma} c_{{\bf{k}}\sigma}
\!+\!\frac{1}{N}\!\sum_{{\bf k}{\bf k}'{\bf q}}
\! V_{{\bf k},{\bf k}'}
c^{\dagger}_{{\bf k} \uparrow}
c^{\dagger}_{{\bf q\!-\!k} \downarrow}
c_{{\bf q\!-\!k}' \downarrow}
c_{{\bf k}' \uparrow}
\end{equation}
where the d-wave separable potential is 
$V_{{\bf k},{\bf k}'} = U_d f({\bf k}) f({\bf k}')$
with $U_d<0$ and $f({\bf{k}})=(\cos k_x - \cos k_y)$.
This potential is a piece of the nearest-neighbour interaction
used in Ref.~\cite{DNM} to model superconductivity in the cuprates.
For simplicity, we study the nearest-neighbor dispersion
$\epsilon_{\bf{k}}=-2t(\cos k_x + \cos k_y)$, with $t=1$, 
on a square lattice with periodic boundary conditions.
The chemical potential $\mu$ is adjusted to obtain the required density $n$.

To investigate the finite temperature
properties of this model in the intermediate coupling regime
($|U_d|$ of order bandwidth), 
we use the ``fluctuation exchange" (FLEX) approximation\cite{FLEX}. 
Since the important correlations in this problem are
in the particle-particle (p-p) channel, we dress the propagator with these
and solve the problem self-consistently. 
Specifically, the vertex (with two incoming and two outgoing legs)
is defined by the standard integral equation written symbolically 
as \cite{MIGDAL}:
\begin{equation}
\Gamma^q_{k,k^\prime}
=
I^q_{k,k^\prime}
\!-\!\tilde I^q_{k,k^\prime}
-
\sum I^q_{k,p}
G(p)
G(q\!\!-\!\!p)
\Gamma^q_{p,k^\prime},
\end{equation}
where $I^q_{k,k^\prime}$ is the p-p irreducible vertex
and $\tilde I$ is the same quantity with twisted outgoing legs. 
Here and below all the quantities are matrices in spin space,
the four-momentum $k=({\bf k},ik_n)$ with $ik_n$ a fermion Matsubara 
frequency ($p$ and $k'$ have the same nature)
and the four-momentum $q=({\bf q},i\nu_m)$ with 
$i\nu_m$ a boson Matsubara frequency, and the symbolic summation means
integrating out (with proper factors) intermediate momenta and frequencies and
matrix multiplication with respect to spin indices.
The Green's function 
$G(k) = [ik_n - \epsilon_{\bf{k}} - \mu - \Sigma(k)]^{-1}$ 
is defined in terms of the self-energy,
which satisfies the relation \cite{MIGDAL}
\begin{equation}
\Sigma(k)
=
\sum V_{k,p}
G(q\!\!-\!\!k)G(p)G(q\!\!-\!\!p)\Gamma^q_{p,k}
\end{equation}
where $V_{k,p}$ is the bare two-body potential.

We now approximate the p-p irreducible vertex by the bare 
potential from (\ref{eq:H}), in which case the contribution from $\tilde I$ 
to $\Sigma$ vanishes. For the self-energy, this is completely equivalent to 
the well-known self-consistent T-matrix approximation \cite{TMATRIX}.
The advantage of the present approach is that in addition to the
one-particle Green's function $\Gamma$ also defines the 
two-particle Green's functions,
and diagrams  with ``twisted'' legs are important for evaluating
response functions such as the spin susceptibility (as we will
discuss in detail elsewhere \cite{FUTURE}).

The coupled integral equations for the self-energy and the vertex part
are numerically solved using fast Fourier transforms\cite{FFT}
and the analytic continuation from Matsubara to real frequencies is 
performed using Pad\'e approximants\cite{PADE}. 
Technical details, as well as checks on the method and the numerics
for the case of the attractive Hubbard model, where we could compare
our results against quantum Monte Carlo \cite{NEGU}, 
will be described elsewhere \cite{FUTURE}.

In this paper we focus primarily on the spectral function 
$A({\bf k},\omega) = -(1/\pi) {\rm Im}G({\bf k},\omega+i0^+)$, 
which is closely related to
the ARPES intensity \cite{ARPES}.
We show representative results for $U_d=-8$ and $n=0.5$
(quarter-filling, $n_\sigma=0.25$); qualitatively similar results were
found for several other parameter sets. All the
results are in the non-superconducting state above $T_c$ (see below). 
In Fig.~\ref{fig:akw:0pi:T2} we plot the spectral functions at a high 
temperature $T=2.0$
for ${\bf k}$ varying from $(0,0)$ to $(\pi,0)$, and see the peaks of
$A({\bf k},\omega)$ disperse through the chemical potential ($\omega = 0$).
We may identify points ${\bf k}^*$ in the Brillouin zone
such that $A({\bf k}^*,\omega)$ has a 
dominant
peak at $\omega=0$.
The ``locus of gapless excitations'' $\{{\bf k}^*\}$ then 
generalizes\cite{FOOTNOTE1} the notion of a 
``Fermi surface'' (FS) to finite temperatures
without any assumptions about well-defined quasiparticles,
and, quite generally, it is a closed contour 
in the repeated-zone scheme. 

It is worth commenting on the lineshapes in Fig.~\ref{fig:akw:0pi:T2}:
at $(0,0)$ the peak is actually infinitely sharp
for our model, and the small width is put in by hand.
Our choice of $V_{{\bf k},{\bf k}'}$ implies 
that states along the diagonal 
$(0,0)$ to $(\pi,\pi)$ are totally unaffected by interactions.
On the other hand, this potential gives rise to very strong
effects near $(\pi,0)$ where the spectral
functions acquire very large widths as seen from Fig.~\ref{fig:akw:0pi:T2}, 
so that
the quasi-particle nature is completely destroyed. 
Note that a log scale is used so that this spreading of spectral
weight over an enormous frequency range can be easily seen.
The sum rule $\int d\omega A({\bf k},\omega) = 1$ is satisfied (to
very high precision) for each ${\bf k}$.

\begin{figure}
  \vspace{-20pt}
  {\epsfxsize=3.30in\epsfbox{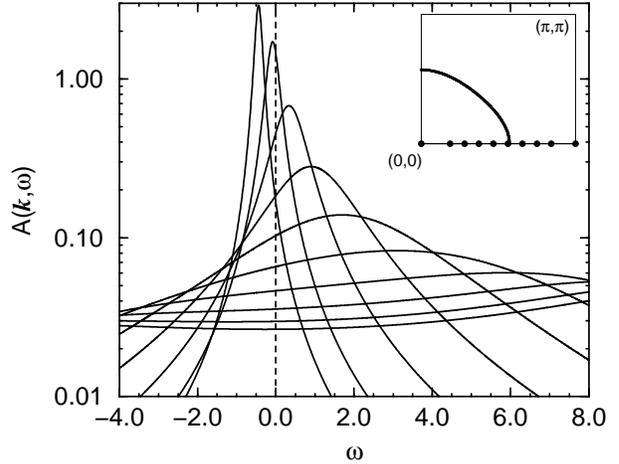}}
  \caption{$A({\bf k},\omega)$ for a sequence of momenta
        ${\bf k}=(x\pi/32,0)$;
        $x=\{0,6,9,12,15,18,21,24,27,32\}$
        for $U_d=-8$ and $n=0.5$ at a high temperature $T=2.0$. 
	In the inset the points indicate the momenta ${\bf k}$ and
	the solid curve shows the $T=0$, non-interacting FS.
        \label{fig:akw:0pi:T2}}
\end{figure}

\begin{figure}
  \vspace{-15pt}
  {\epsfxsize=3.30in\epsfbox{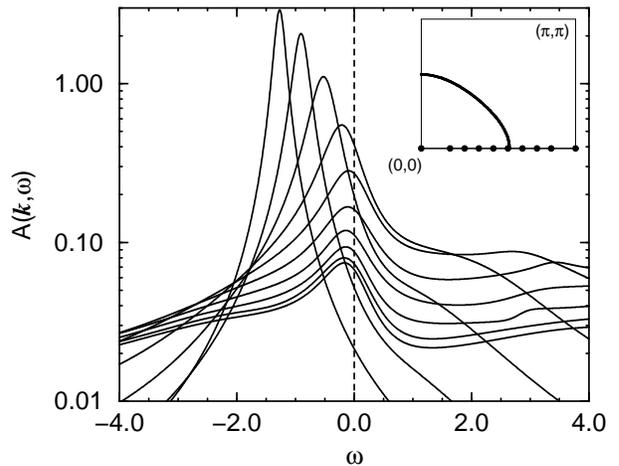}}
  \caption{$A({\bf k},\omega)$ for a sequence of momenta 
       ${\bf k}=(x\pi/32,0)$;
       $x=\{0,6,9,12,15,18,21,24,27,32\}$
        for $U_d=-8$ and $n=0.5$ at $T=0.2$. 
        \label{fig:akw:0pi}}
\end{figure}

As the temperature is lowered below a scale $T^{*} \simeq 1$
(for the same choice of parameters as in Fig.~\ref{fig:akw:0pi:T2}), remarkable
changes takes place in the spectra, as seen from Fig.~\ref{fig:akw:0pi},
which shows results at $T=0.2$ (but still above $T_c$).
First, we find that spectra which showed one broad feature at
high $T$ now show a multiple peak structure (see further below).
Second, the dominant peaks of $A({\bf k},\omega)$ exhibit 
very anomalous dispersion: as ${\bf k}$ varies from $(0,0)$ to $(\pi,0)$, 
the peak approaches $\omega=0$
but never crosses it, either ``bouncing back'' towards negative $\omega$
or ``stalling'' (depending on parameter values),
in complete contrast to the high $T$ results described above.
This is exactly like the pseudogap behavior seen in 
ARPES experiments \cite{STANFORD,ARGONNE}
on underdoped cuprates in the temperature regime $T_c < T < T^*$.

To understand the multiple peak structure of the spectral functions
it is useful to look at plots of the real and imaginary parts of the
(retarded) self-energy as functions of $\omega$. 
We find that the dominant peak of $A({\bf k},\omega)$, 
which is at $\omega \le 0$ for each ${\bf k}$ in Fig.~\ref{fig:akw:0pi}, 
is associated with a solution of
$\omega - \epsilon_{\bf k} - \mu - {\rm Re}\Sigma({\bf k},\omega) = 0$
for which $d{\rm Re}\Sigma/d\omega < 0$ and ${\rm Im}\Sigma$ is small.
The very broad features at $\omega > 0$ come from solutions for which
${\rm Re}\Sigma$ has a positive slope and ${\rm Im}\Sigma$ is large.
It is tempting to describe the ``bounce'' and multiple-peak structure
as a precursor of the Bogoliubov-like dispersion of excitations in the SC
state. In the presence of strong self-energy effects, 
such a simple picture may need to be generalized significantly.

\begin{figure}
  \centerline{\epsfxsize=1.20in\epsfbox{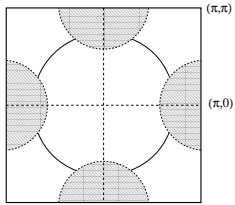}}
  \vspace{5pt}
  \caption{Fermi Surface destroyed in patches;
  solid lines represent gapless excitations and the shaded patches
  indicate momenta where there is strong scattering and the FS is washed out.
  \label{fig:FS}}
\end{figure}

Let us now ask how the pseudogap affects the ``Fermi surface'' (FS) 
by studying the ``locus of gapless excitations'' 
$\{{\bf k}^*\}$ (defined, as before, by the condition that
the spectral function at that ${\bf k}$ has a dominant peak centered at
$\omega = 0$). Along (or near) the zone diagonal,
$(0,0)$ to $(\pi,\pi)$, interaction effects are absent (or weak) and
there is a conventional FS crossing with a well defined ${\bf k}^*$. 
However, along the $(0,0)$ to $(\pi,0)$ there is no FS crossing.
By studying the spectral function in the entire zone we find that
the anomalous dispersion and large line-widths
destroy the notion of a FS as a closed contour
of gapless excitations (even in the repeated-zone scheme).
The resulting picture emerging from our calculations,
and consistent with ARPES experiments \cite{STANFORD,ARGONNE,SHEN-JRS}, 
is that the Fermi Surface is destroyed in patches in the
Brillouin zone, as schematically depicted in Fig.~\ref{fig:FS}.

\begin{figure}
  \vspace{-15pt}
  {\epsfxsize=3.30in\epsfbox{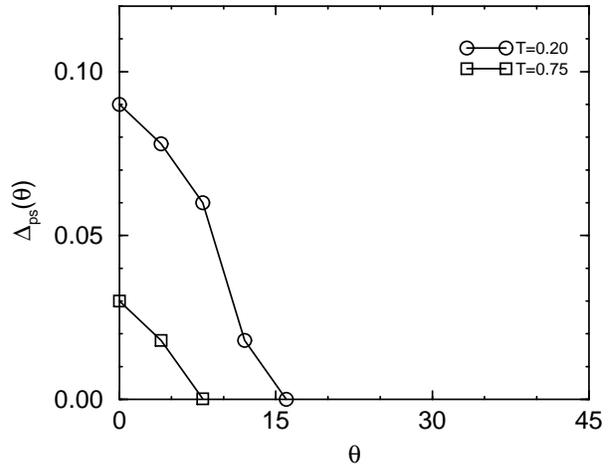}}
  \caption{The angle dependence of the $\Delta_{ps}(\theta)$ 
           for temperatures $T=0.2$ and  $T=0.75$ 
           and $U_d=-8$, $n=0.5$.
           \label{fig:pseudo:T}}
\end{figure}

A more quantitative understanding of the destruction of the FS
can be obtained by studying the angular dependence of the
pseudogap and its variation with temperature.
In the spirit of the ARPES experiments
we estimate the pseudogap by making scans through
${\bf k}$-space and noting the position of the spectral function
peak which is farthest to the right, i.e. at the largest frequency
below zero. We then plot the spectral function pseudogap
$\Delta_{ps}$ as a function of $\theta=\arctan(k_y/k_x)$.
In Fig.~\ref{fig:pseudo:T} we plot the angle-dependence of 
$\Delta_{ps}(\theta)$ at two temperatures $T=0.2t$ and $T=0.75t$.
The first point to notice is the strong anisotropy of the gap,
which is always suppressed to zero in an arc about the diagonal.
The extent of the zero gap (nodal) region , and the magnitude
of the maximum gap at low $T$ are both sensitive functions of the choice
of parameters (see below); we find that the larger the maximum gap, the smaller
the nodal region. 
The second important point to note is the $T$-dependence of the pseudogap,
which suggests a gap collapse due to quasiparticles excited 
around the nodal regions in a d-wave SC gap, similar to a 
recent suggestion of Lee and Wen \cite{LEE-WEN}.
At high temperatures, the pseudogap gets suppressed and eventually disappears 
upon further heating; above a crossover scale $T^* \simeq 1$ 
(for the parameters discussed here) the spectral function peaks disperse 
through $\omega=0$ for all fixed-$k_y$ scans.
Thus we find that we recover a closed contour of gapless excitations
above $T^*$. However, in our model we find that this ``Fermi surface''
has $T$-dependence \cite{FOOTNOTE1}.
It is quite remarkable that in the ARPES experiments, there are 
indications\cite{ARGONNE} of an underlying Luttinger FS which 
is $T$-independent within error bars. 

\begin{figure}
  \vspace{-15pt}
  {\epsfxsize=3.30in\epsfbox{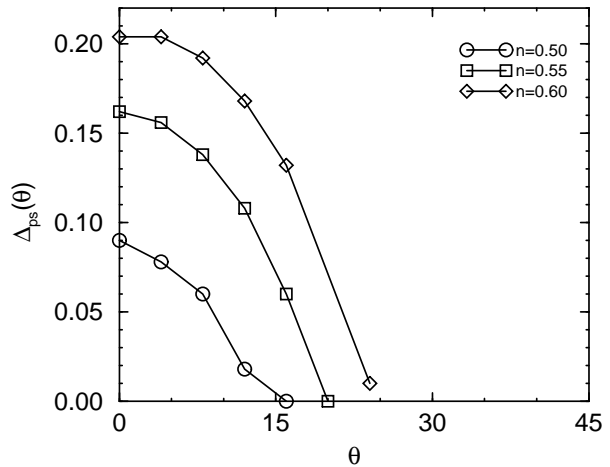}}
  \caption{The angle dependence of $\Delta_{ps}(\theta)$
           for densities $n=0.5$, $n=0.55$  and  $n=0.60$ 
           and $U_d=-8$, $T=0.2$.
           \label{fig:pseudo:n}}
\end{figure}

It is instructive to consider the doping dependence of the effect.
In Fig~\ref{fig:pseudo:n} we show $\Delta_{ps}(\theta)$ at the same 
temperature and coupling, for three values of the density.
The pseudogap clearly increases rapidly with increasing density.

\begin{figure}
  \vspace{-15pt}
  {\epsfxsize=3.30in\epsfbox{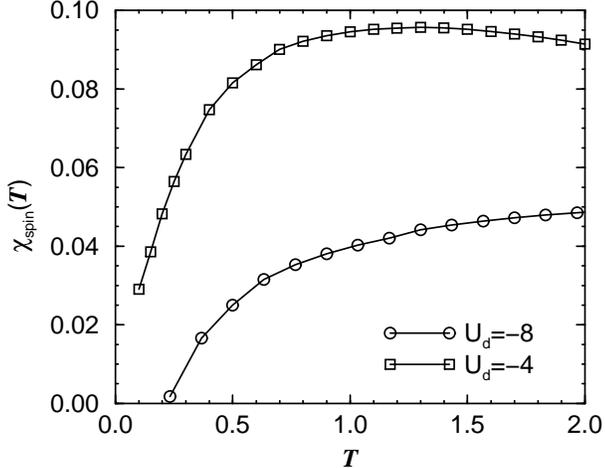}}
  \caption{The temperature dependence of the spin susceptibility
           for $U_d=-8$ and $U_d=-4$
           at density $n=0.5$.
           \label{fig:spin:chi}}
\end{figure}

Up to now, we have discussed the pseudogap behaviour in our model in
the single-particle spectral weight.  In Fig.~\ref{fig:spin:chi} we show
that the temperature dependence of the static, uniform,
spin susceptibility
also has a spin gap, similar to what is observed in NMR\cite{NMR} and
has been calculated\cite{NEGU} for the case of the attractive Hubbard model
which has s-wave pairing.
More details of this calculation will be presented
elsewhere \cite{FUTURE}.

To ensure that we are {\it above} the superconducting $T_c$
we calculate the parameter
$\lambda=|U_d|\chi^0_{pp}(q=0)$ where
$\chi^0_{pp}(q)=(T/N)\sum_{k} f^2(k)G(k)G(q-k)$.
The gap equation at $T=T_c$ reduces to the
condition $\lambda=1$, which in our
approximation coincides with the appearance of a pole in the full vertex.
Thus, one should keep $\lambda<1$ to be in the normal state.
In 2D, there is of course no LRO at non-zero temperatures and one
expects $\lambda<1$ for all $T>0$.  Even though our formalism does
not include topological excitations, it is reasonable to expect
$\lambda\approx1$ near the onset of algebraic order.
In practice we consider only temperatures for which $\lambda<0.8$.

We conclude this paper by showing that the 
pseudogap behavior described above is occuring in 
{\it a degenerate Fermi system above} $T_c$. 
This is important to establish since gap-like features can be
trivially obtained either in a system below $T_c$, or in a strongly
coupled regime where all the electrons are tightly bound up 
into bosonic pairs.
The specific heat\cite{SP-HEAT} and the ARPES\cite{STANFORD,ARGONNE}
experiments clearly indicate that in the underdoped systems one is still
dealing with a degenerate Fermi system and {\it not} bosons.
By looking at the momentum distribution $n({\bf k})$ \cite{FUTURE}, 
and by ascertaining that the chemical potential $\mu \gg T$, we know
that we are in a degenerate Fermi regime.


In this paper we have studied the normal state of a simple model
in which d-wave pairing correlations above $T_c$
lead to the appearance of a highly anisotropic pseudogap and
the destruction of the Fermi surface, which are remarkably
similar to the results of ARPES experiments. Quantitative comparison
with the experiments must, however, await a controlled
calculation based on a microscopic model
which describes how a Mott insulator upon doping
goes into a short coherence d-wave superconductor whose
normal state is dominated by pairing correlations.

A.~N. and J.~E. are supported by Boston College.
Additional support by the National High Magnetic Field
Lab and Los-Alamos National Lab is acknowledged.
J.~E. and M.~R. would like to thank Professor Yu Lu and the 
ICTP, Trieste, for hospitality during a visit which helped
to initiate this research.


\begin{references}
\bibitem{FFT} The FFT works reliably with functions that decay as
$1/\omega^2$ (or faster) at large $\omega$ because of
the necessity of a cutoff. For the present model 
a cutoff 10 to 20 times the bandwidth was necessary.
Our runs were typically for 64$\times$64 lattices with a 
uniform $\omega$ mesh of 512 points.
\bibitem{FOOTNOTE1}
The gapless locus for a strongly correlated system at finite temperatures, 
need not posses all of the properties that we associate
with the usual $T=0$ definition of a FS in a Fermi liquid.
For instance, there is no guarantee that the volume enclosed by this
locus is $T$-independent. Nevertheless, by focusing on gapless excitations,
it seems to be a very natural generalization.
\end{references}
\end{document}